\documentclass[aps,prl,reprint,groupedaddress,footinbib,showpacs]{revtex4-1}

\usepackage{graphicx}
\usepackage{upgreek} 
\usepackage{amsmath} 

\usepackage{pifont}
\usepackage{lipsum}
\usepackage{color} 

\newcounter{mycounter}

\begin{document}


\title{Rydberg polaritons in a thermal vapor}


\author{Fabian Ripka}
\email[Electronic address: ]{f.ripka@physik.uni-stuttgart.de}
\affiliation{5. Physikalisches Institut and Center for Integrated
Quantum Science and Technology, Universit\"at Stuttgart,
Pfaffenwaldring 57, 70569, Stuttgart, Germany \\}
\author{Yi-Hsin Chen}
\email[Electronic address: ]{yhchen920@gmail.com}
\affiliation{5. Physikalisches Institut and Center for Integrated
Quantum Science and Technology, Universit\"at Stuttgart,
Pfaffenwaldring 57, 70569, Stuttgart, Germany \\}
\author{Robert L\"{o}w}
\affiliation{5. Physikalisches Institut and Center for Integrated
Quantum Science and Technology, Universit\"at Stuttgart,
Pfaffenwaldring 57, 70569, Stuttgart, Germany \\}
\author{Tilman Pfau}
\email[Electronic address: ]{t.pfau@physik.uni-stuttgart.de}
\affiliation{5. Physikalisches Institut and Center for Integrated
Quantum Science and Technology, Universit\"at Stuttgart,
Pfaffenwaldring 57, 70569, Stuttgart, Germany \\}



\date{\today}

\begin{abstract}
We present a pulsed four-wave mixing (FWM) scheme via a Rydberg state to create, store and retrieve collective Rydberg polaritons. The storage medium consists of a gas of thermal Rb atoms confined in a 220\,$\mu$m thick cell, which are heated above room temperature. The experimental sequence consists of a pulsed excitation of Rydberg polaritons via the D1 line, a variable delay or storage time, and a final retrieval pulse via the D2 line.
The lifetime of the Rydberg polaritons is around 1.2\,ns, almost entirely limited by the excitation bandwidth and the corresponding motional dephasing of the atoms.
The presented scheme combined with a tightly confined atomic ensemble is a good candidate for a deterministic single-photon source, as soon as strong interactions in terms of a Rydberg blockade are added.
\end{abstract}


\maketitle
Photons make very good flying quantum bits (qubits) because of their weak interaction with the environment. The ability to store a qubit is essential for long-distance quantum communication~[\citenum{Chaneliere2005,Eisaman2005,DLCZ2001}]. This works best if the optical properties of the medium match the bandwidth of the qubits. Scientists have achieved a high efficiency quantum memory as well as a single-photon-level memory in a $\Lambda$-type atomic transition~[\citenum{YHChen2013,Hosseini2011,Reim2011}]. The quantum information can be encoded in a polariton~[\citenum{Fleischhauer2000}], which is in a more general sense a coherent superposition of light and matter. If the transfer to a purely matter-like excitation is coherent and reversible, the quantum information can be mapped back onto a light field.
To implement in addition quantum information processing, strong interactions between stored qubits are required.
The atomic excitation involving a high-lying Rydberg state matches these requirements, as it allows for storage, control and retrieval of optical photons via Rydberg polaritons~[\citenum{Maxwell2013}].
It also provides strong long-range interactions and blockade effects~[\citenum{Saffman2010,Gorshkov2011}], which enables novel quantum devices, such as single-photon transistors~[\citenum{Gorniaczyk2014,Tiarks2014}], quantum gates~[\citenum{Jaksch2000,Wilk2010,Isenhower2010}], and deterministic single-photon sources~[\citenum{Dudin2012}]. Nevertheless extremely tight traps and low temperatures are required to achieve such Rydberg based non-linearities. If such quantum devices can also be realized with thermal atoms, scalability and integrability is within reach.

Rydberg-EIT (electromagnetically induced transparency) in thermal vapors has first been studied in Ref.~[\citenum{Mohapatra2007}] without any notion of an Rydberg-Rydberg interaction. For the direct observation of interacting Rydberg atoms at room temperature, a bandwidth-limited pulsed excitation was applied to determine interaction strengths corresponding to a few GHz, much stronger than the decoherence rate set by the motion-induced Doppler effect~[\citenum{Huber2011,Baluktsian2013}].
Many-body phenomena based on this interaction like aggregation~[\citenum{Lesanovsky2014,Urvoy2015}] or optical bistabilities~[\citenum{Carr2013}] have been studied in thermal vapor cells. If one adds a de-exciation laser before the Rydberg states do some unwanted stuff like aggregation etc., the corresponding FWM scheme provides the basis for a deterministic single-photon source~[\citenum{Muller2013}]. It is important that the polariton lifetieme, or better said the coherence time, is sufficient to produce a mixed wave fully coherent.
The Rydberg polariton lifetime can be limited by dephasing, e.g. due to inhomogeneous differential light shifts in an optical dipole trap for cold atoms~[\citenum{Li2013}]. In thermal vapors, dipole traps are not applicable but the lifetime is now limited by motional dephasing. The corresponding bandwidth determines the blockade radius for a given van der Waals interaction.

\begin{figure*}[t]
\centering
\includegraphics[width=1.8\columnwidth]{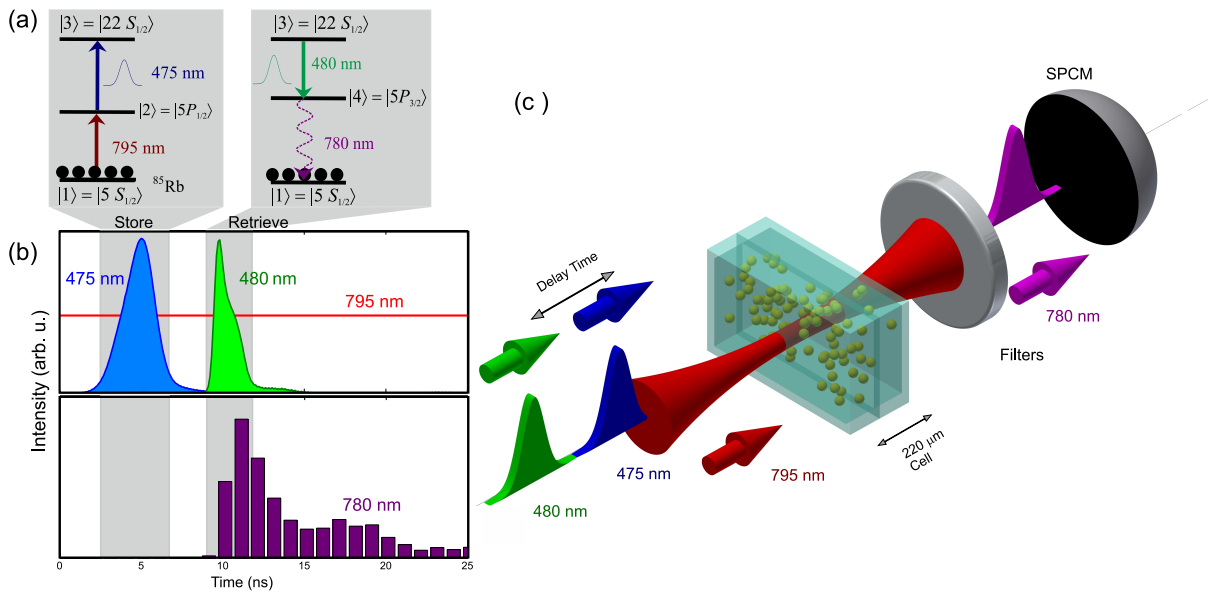}
\caption{(Color online) Principle of the experiment.
(a) Relevant energy levels and store-and-retrieve process: first a pulsed two-photon excitation to the Rydberg state $|22 S\rangle$, then free evolution of the coherence, followed by the retrieval to the intermediate state $|4\rangle$, from where the four-wave mixing process is closed by emitting photons at $\sim$780\,nm.
(b) Time sequence and intensity profiles of the generated light and applied light fields.
(c) Simplified scheme of the experimental setup. All laser beams are co-propagating through a 220 $\mu$m thick cell. Four bandpass filters are inserted in front of a SPCM detector to suppress the three incident fields. An additional neutral-density (ND) filter is used to attenuate the generated signals to the single-photon level.
\label{fig:setup}}
\end{figure*}

In this Letter, we investigate the properties of thermal Rydberg polaritons in the absence of Rydberg-Rydberg interactions by a double-pulsed FWM transition. Some of the incident photons are first stored as matter-like polaritons. A pulsed retrieval laser converts the excitations back into light-like polaritons, i.e. photons which we detected. The retrieval laser is temporally delayed in order to probe the effective lifetime of the polaritons.
We present measurements for different storage times of Rydberg polaritons, and obtain the lifetime of polaritons to be 1.2\,ns for our given bandwidth of the excitation sequence. We detect the retrieved photons produced via FWM by single-photon counters as we only address quite small atomic ensemble.
The retrieval light field shows motion-induced revivals, which can be attributed to the constructive interference between the coherent evolution of different velocity classes~[\citenum{Huber2014,YHChen2015}]. All these features can be reproduced by numerical simulations considering the contribution of a whole Doppler ensemble. It is important to note that the temporal evolution can only be explained by a collective response of the ensemble and not by a single-atom master equation. Such a collective behavior is also necessary for the realization of a single photon source as detailed in ~[\citenum{Muller2013}]. Rydberg interactions do not play a role here due to the low atomic density and low principal quantum number ($|22S\rangle$). However, our research paves the way for the studies involving Rydberg interactions and coherent control of the stored Rydberg polaritons.


The experiment is carried out in a $\sim$\,220 $\rm\mu$m glass cell filled with rubidium consisting of the two stable isotopes at natural abundance. The atoms are heated above room temperature to T $\sim 140^\circ \mathrm{C}$.
The measured atomic density in the ground state $|5S_{1/2},F=3\rangle\equiv|1\rangle$ of $^{85}$Rb is around $3.0 (\pm 0.8) \times10^{12}\mathrm{cm}^{-3}$.
A resonant cw 795\,nm laser reduces the atomic density to be $1.3 (\pm 0.4)\times10^{12}\mathrm{cm}^{-3}$ by optically pumping the populations into the other hyperfine ground state.
The schematic experimental timing and setup are shown in Fig.~\ref{fig:setup}.
The Rydberg polaritons are created by a two-photon transition with the cw 795\,nm laser and a bandwidth-limited pulsed 475\,nm laser of FWHM $\sim$ 2.5 ns, driving  a resonant transition to the Rydberg state $|22S\rangle \equiv |3\rangle$ via the intermediate state $|5P_{1/2},F'=3\rangle \equiv |2\rangle$.
After a time delay, a pulsed retrieval field with wavelength of $\sim$ 480 nm imprints the coherence onto state $|4\rangle = |5P_{3/2}, F'=4\rangle$, generating the FWM signals at $\sim$ 780 nm.
The temporal delay between the two pulses is set by a variable optical delay line, similar to Ref.~[\citenum{Klovekorn1998}].
In order to detect the retrieved signals, four additional bandpass filters are inserted in front of a single-photon counting module (SPCM) to suppress the three incident fields. An additional neutral-density (ND) filter is used to attenuate the generated signals to the single-photon level. The repetition rate is 50 Hz.
All lasers are linearly polarized in the same direction and co-propagating through the vapor cell.
The 795 nm laser is focused onto the cell with a $1/e^2$ beam diameter of around $35(5)$ $\mu$m. Both blue beams are overlapped by a pinhole which is imaged into the cell to provide a homogeneous beam profile among the interaction regime at a diameter of $\sim 67(5)$ $\mu $m. We determine the peak Rabi frequencies applied in this experiment to be $\Omega_{795}/2\pi = 85(10)$ MHz, $\Omega_{475}/2\pi = 160(10)$ MHz, and $\Omega_{480}/2\pi = 130(10)$ MHz. More details can be found in Ref.~[\citenum{setup}].

The time-evolution of the atom-light system can be calculated by a basic four-level transition model coupled by three light fields ~[\citenum{Huber2014}]. We numerically solve the Lindblad equation with the Hamilton operator in a four-level diamond configuration, $\frac{\partial \hat{\rho_v}}{\partial t}=-\frac{i}{\hbar }\left[\hat{H_v},\hat{\rho_v}\right] + L (\hat{\rho_v})$ for the velocity group $v$. The Lindblad operator $L (\hat{\rho_v})$ includes the various decays~[\citenum{Kolle2012}].
The corresponding Hamiltonian with rotating-wave approximation is
\begin{equation}
\hat{H_v}=\hbar  \left(
\begin{array}{cccc}
 0 & \frac{\Omega _{795}(r)}{2} & 0 & 0 \\
 \frac{\Omega _{795}^*(r)}{2} & \delta_{v,795} & \frac{\Omega _{475}(r,t)}{2} & 0 \\
 0 & \frac{\Omega _{475}^*(r,t)}{2} & \delta_{v,Ryd} & \frac{\Omega _{480}(r,t)}{2} \\
 0 & 0 & \frac{\Omega _{480}^*(r,t)}{2} & \delta_{v,480} \\
\end{array}
\right).
\end{equation}
$\delta_{v,795}$, $\delta_{v,Ryd}$, and $\delta_{v,480}$ are the respective Doppler detunings.
The spontaneous decay rates from intermediate states are $\rm \Gamma_{2}/2\pi=5.7$\,MHz and $\rm \Gamma_{4}/2\pi=6.0$\,MHz (from state $|2\rangle$ and $|4\rangle$ respectively); from the Rydberg state $|22S\rangle$ to each intermediate state they are $\rm \Gamma_{42}/2\pi=\rm \Gamma_{43}/2\pi=23$\,kHz and transient effects due to the atomic motion give rise to effective damping of $\Gamma_t/2\pi = 2.47$\,MHz.

The temporal structure of the retrieved signals is sensitive to the incident laser intensities in a resonant FWM transition~[\citenum{YHChen2015}]. Hence, we further consider the spatial profiles of the beams. The 795\,nm laser beam has a Gaussian profile with beam diameter $2w =35\,\mu$m at focus, so that the Rabi frequency of the 795\,nm laser is
$\Omega _{795}(r)=\Omega _{795,\text{peak}}e^{-r^2/{w^2}}$, where $\Omega_{795,\text{peak}}$ is the peak Rabi frequency. We assume that the beams at 475\,nm and 480\,nm have homogeneous profiles in the excitation regime. In addition, the Gaussian atomic velocity distribution is divided into linearly discrete intervals with 200 velocity classes, up to $|v|$ = 500 m/s. Summing up the coherence $\rho_{41}$ from different velocity classes with a Gaussian weight, we obtain the collective coherence at focus $\rho _{41}\left(r,t\right)={\int \rho _{41}\left(v,r,t\right)\times e^{-\frac{mv^2}{2 k\text{T}}}dv}/\int e^{-\frac{mv^2}{2k\text{T}}} \, dv$,
where $m$ is the atomic mass and $k$ is the Boltzmann constant.

The connection of the coherence and the electric field can be derived by the wave equation of motion with slowly varying amplitude approximation. We get

\begin{equation}
\begin{array}{r}
\frac{\partial }{\partial z}E_0(z,r,t)=-\frac{i\omega}{2 c \epsilon _0}P_0(z,r,t) \\
= -\frac{in\omega}{2 c \epsilon _0 }d_{41}\rho _{41}(z,r,t).
\end{array}
\end{equation}
$E_0$ and $P_0$ are the amplitudes of the electric field and the polarization, respectively; $n$ is the atomic density in the ground state; $d_{41}$ is the dipole matrix element of states $|1\rangle$ and $|4\rangle$, which is $1.67\times 10^{-29}$ C m; and $\omega$ is the angular frequency of the generated 780\,nm light field. We assume the amplitude value of the electric field to be $z$-independent since the Rayleigh range of the focused beam is much larger than the length of the cell $L$. After propagating through the atomic ensemble, we derive the generated electric field right behind the cell to be
\begin{equation}
E_{0,\rm cell}(r,t)=-\frac{in\omega }{2 c \epsilon _0}d_{41}\rho _{41}(r,t)L.
\end{equation}

The generated electric field can be filtered out by the serial optical detection system. Taking into account the apertures at the imaging system including two lenses and one pinhole~[\citenum{propagation}], we derive the detected electric field after pinhole $E_{\rm out}(r,t)$ by applying the Fresnel approximation and phase transformation function right behind lenses $E(r,t)e^{-ik_{780}r^2/{2 f}}$ ($f$ is the focal length and $k_{780}$ is the wave-vector of the signal light)~[\citenum{Goodman}]. By this method, we get the time-dependent power $P(t)$ and photon number $N_{\rm ph}$ of the generated light,
\begin{align}
P(t) &= \frac{c \epsilon _0}{2} \int_{r < r_{\rm hole}} |E_{\rm out}(r,t)|^2 2 \pi  r dr,\nonumber\\
N_{\rm ph} &= \int P(t) \, dt/\hbar \omega.
\end{align}
$r_{\rm hole} = 5\,\mu$m is the radius of the pinhole.


\begin{figure}[t]
\centering
\includegraphics[width=0.9\columnwidth]{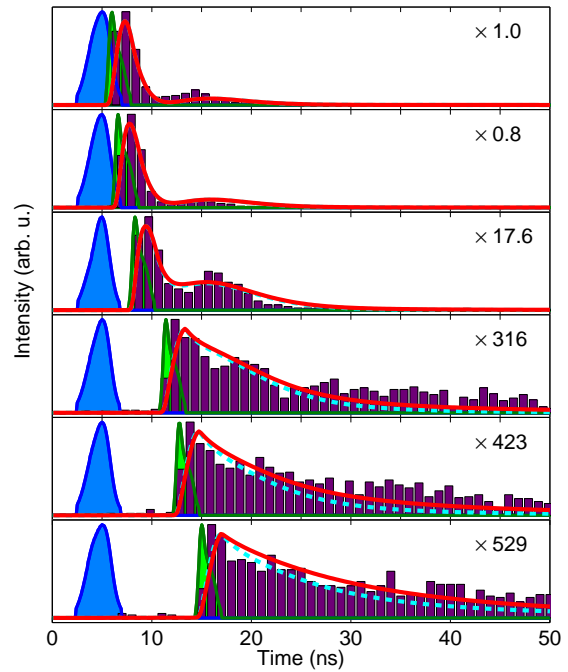}
\caption{(Color online) The retrieved signal dynamics for different storage times. The temporal signals are depicted as histograms. The pulse shapes are shown in blue and green for the storage field and the retrieval field, respectively. The results are normalized to the signal in the top plot and that further times the values shown in the each subplot. The cyan dashed lines are the simulation results of coherent emission; and the red solid lines are the results further including spontaneous emission. $\Omega_{795, \rm peak}/2\pi $, $\Omega_{475}/2\pi$, and $\Omega_{480}/2\pi$ are set as 70, 160, and 130 MHz for the simulation.
\label{fig:dynamics}}
\end{figure}

The experiment begins with a two-photon excitation. Few photons are first stored in the medium as Rydberg polaritons after the pulse is over. The coherence $\rho_{31}$ is then imprinted onto the D$_2$ transition by applying a retrieval pulsed laser at 480\,nm after a variable storage time.
We first measure the bandwidth of the coherence $\rho_{31}$ by systematically scanning the detuning of the retrieval laser.
The storage time is fixed at 3.1\,ns. The measured full width at half maximum (FWHM) bandwidth is $\sim$ 500 MHz, which is the convolution of the $\rho_{31}$ lineshape and the spectrum of the retrieval pulse.
In the following measurements, the 480\,nm laser is locked at resonance frequency.

We then vary the storage times of the Rydberg polaritons. Looking at the dynamics of the retrieved signals measured at short delay times (Fig.~\ref{fig:dynamics}), we observe motion-induced signal revivals, as discussed in Refs.~[\citenum{Huber2014,YHChen2015}]. The atomic coherences differ in both amplitude and phase for the different velocity classes. After the excitation pulse, the atomic coherences are evolving freely according to their Doppler detunings with phase $\propto$ exp(-$i k_{780}vt$). The temporal shape of the oscillatory FWM signals can be attributed to the coherence interference of different atomic velocity classes.
The cyan dashed lines are the numerical results from solving the master equation and extracting the coherence $\rho_{41}$.
For longer storage times, we observe slowly decaying tails of the retrieved signals.
We attribute this to the effect of spontaneous emission which in the simulation is derived from $\Gamma_4 \cdot \rho_{44} \times$ collection efficiency~[\citenum{propagation}].
The numerical results including both types of emission, shown as the red solid lines, fit the measurements well.

\begin{figure}[t]
\centering
\includegraphics[width=.95\columnwidth]{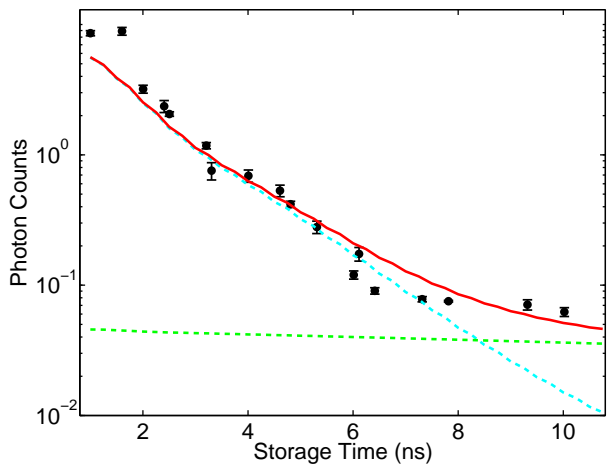}
\caption{(Color online)
Dependence of the 780\,nm photon counts per pulse pair on the temporal delay (storage time) of the pulses. Note that the storage time shown here represents the time difference of the peaks of the storage and retrieval pulses.
The cyan and green dashed lines are the simulation of coherent and spontaneous emissions, respectively, with the parameters used in Fig.~\ref{fig:dynamics}. The red solid line is the sum of both types of emission. The error bars represent one standard deviation based on the measurement statistics.
\label{fig:lifetime}}
\end{figure}

With 1.0\,ns storage time, 9\, retrieved photons are detected. For shorter storage times, the signals are dominated by the coherent collective emission; and for longer delay times, the signals are mainly dominated by the spontaneous emission. The cyan and green dashed lines shown in Fig.~\ref{fig:lifetime} represent the simulation of the coherent and spontaneous emissions, respectively, with the parameters used in Fig.\ref{fig:dynamics}, and the red solid line is the sum of both types of emission. The adjusted density in the simulations is $0.92 \times10^{12}\mathrm{cm}^{-3}$, which is in the range of the experimentally determined density.
An exponential fit to the coherent part of the emission (i.e. for storage times less than 6.5\,ns) shows a polariton lifetime of 1.2\,ns.
The lifetime is not dominated by the Rydberg population decay of 6.9\,$\mu$s neither by the transient time (the time period that atoms move in and out of the interaction regime is around 64\,ns). The lifetime is inferred from the Doppler dephasing.
Due to the co-propagating configuration of the excitation laser beams, the effective wave vector to the Rydberg state is $k_{\rm eff} = k_{795} + k_{475}$.
The FWHM bandwidth of Rydberg population is $\rm BW_{coh}$$= k_{\rm eff}\cdot\Delta v$. From the numerical calculation, we know that the velocity band of Rydberg-excited atoms has a FWHM of $\Delta v = 52$\,m/s, implying the bandwidth of $\rm BW_{coh}/2\pi = 175$ MHz. The corresponding Doppler dephasing lifetime is 0.91\,ns, close to the measurements.

In future works, we will increase the Rydberg quantum number to 37 and above, for example, for which interaction energy of up to $\sim$ 2 GHz~[\citenum{Baluktsian2013}] is much larger than the measured excitation bandwidth.
Moreover, for these parameters a Rydberg blockade volume of $\sim 6~\mu m^3$ is expected which can be matched by reducing the excitation volume by combining a tightly focused beam and a micrometer-sized cell. Therefore, this scheme can be used for generating non-classical light fields~[\citenum{Muller2013}].


In summary, we store and retrieve thermal Rydberg polaritons in a bandwidth-limited double-pulsed scheme. With 1.0\,ns delay time, we obtain 9 retrieved photons. The thermal Rydberg polaritons have a lifetime of 1.2\,ns, almost entirely limited by motional dephasing of the atoms. Our numerical calculation including the spatial profiles of the laser beams and also the light propagation of the generated field fits well to the temporal structures and lifetime of the retrieved signals. We propose that this FWM-excitation scheme combined with strong Rydberg interactions can be used for generating non-classical light fields or manipulating quantum information carriers by photons.

\begin{acknowledgments}
This work is supported by the ERC under Contract No. 267100, BMBF within Q.com-Q under Project No. 16KISO129. Y.H.C. acknowledges support from the Alexander von Humboldt Foundation.
\end{acknowledgments}

\bibliographystyle{apsrev4-1}   
\bibliography{RydbergReference}   

\begin{thebibliography}{31}
\bibitem [{set()}]{setup}%
  \BibitemOpen
  \href@noop {} {\bibinfo  {journal} {These two pulsed laser fields are
  generated by two dye amplifier systems, one is a self-built quad-pass setup
  and one is a commercial setup. Both dye amplifiers are pumped by the same
  pulsed Nd: YAG laser. The powers of all laser beams have 20$\%$ reduction
  after passing through the uncoated cell. The measured data in Fig. 2 are
  averaged up to 6,000 runs. The error bars in Fig. 3 represent one standard
  deviation extract from 5 measurements, each averaged up to 1,300 runs}\
  }\BibitemShut {NoStop}%
\bibitem [{pro()}]{propagation}%
  \BibitemOpen
  \href@noop {} {\bibinfo  {journal} {The detection system includes one 25
  mm-diameter lens with focal length 100 mm, one 4 mm-diameter lens with focal
  length 18.4 mm, and one pinhole with diameter of 10 $\mu$m. The collection
  efficiency of the signals by a multimode fiber and SPCM detector is around
  $40\%$. The collection efficiency of the spontaneous emission is set as
  $1.1\times 10^{-5}$, which is close to the derived factor by considering the
  solid angle and the propagation loss}\ }\BibitemShut {NoStop}%
\end{thebibliography}


\end{document}